\documentclass[aps,prb,twocolumn,superscriptaddress,amsmath,nofootinbib]{revtex4-1}
\usepackage{graphicx}
\begin{document}
\title{Enhanced energy harvesting from shadow-effect: mechanism and a new device geometry}
\author{Amit K. Das}
\email[Corresponding author e-mail: ]{amitdas@rrcat.gov.in}
\affiliation{Oxide Nano-electronics Laboratory, Laser Material Processing Division, Raja Ramanna Centre for Advanced Technology, Indore 452013, India}
\author{V. K. Sahu}
\affiliation{Oxide Nano-electronics Laboratory, Laser Material Processing Division, Raja Ramanna Centre for Advanced Technology, Indore 452013, India}
\affiliation{Homi Bhabha National Institute, BARC Training School Complex, Anushaktinagar, Mumbai- 400 094, India}
\author{R. S. Ajimsha}
\affiliation{Oxide Nano-electronics Laboratory, Laser Material Processing Division, Raja Ramanna Centre for Advanced Technology, Indore 452013, India}
\author{P. Misra}
\affiliation{Oxide Nano-electronics Laboratory, Laser Material Processing Division, Raja Ramanna Centre for Advanced Technology, Indore 452013, India}
\affiliation{Homi Bhabha National Institute, BARC Training School Complex, Anushaktinagar, Mumbai- 400 094, India}

\date{\today}

\begin{abstract}
Energy harvesting from shadow-effect is the generation of electrical power from a Schottky junction when a part of it is kept in shadow and the remaining under illumination. It has been recently invented in Au/n-Si junctions, where modulation of work function of the Au top electrode under contrasting illumination has been invoked to explain the effect. In this communication, a completely different physical mechanism for energy harvesting from shadow-effect in a Schottky junction is proposed that does not assume change in work function of the top electrode under illumination. The device, termed shadow-effect energy generator (SEG), is modelled as two parallel Schottky junction solar cells, one at the shadowed and the other at the illuminated part, connected with each other in a closed loop circuit through the Si substrate and the top electrode. To test the proposed mechanism, ITO/n-Si junction based SEGs have been fabricated. The values of open circuit voltage in the SEGs have been found to be matching with the difference of photovoltages of the two cells corrected for the potential drop across the Si substrate, that validates the proposed mechanism. To further corroborate the mechanism, the conventional SEG geometry has been modified by applying a continuous ohmic coating at the back of the Si substrates that bypasses the resistance of the Si substrate for current flow and results in higher open circuit voltage and short circuit current.  Moreover, ITO/n-Si based SEGs have been found to produce higher output power density compared to that reported in Au/n-Si devices in both the conventional and the new geometry. Although the closed loop present in the equivalent circuit of the SEG devices lead to wastage of harvested energy, the ITO/n-Si SEG devices can nevertheless be used as self-powered sensor for light, object and movement detection as well as for producing electricity from contrasting illumination.
\end{abstract}
\maketitle

\section{Introduction}
Energy harvesting from shadow effect is a phrase coined to describe a recently reported method to generate electricity from contrast of light and shadow. This phenomenon has been shown in Au/n-Si Schottky junctions, \cite{1} wherein a part of the gold electrode is kept in shadow while the remaining in illumination. A potential difference is generated between the illuminated and the shadowed part of gold electrode which could drive current through external circuit. Work function modulation of the Au electrode under illumination has been invoked to explain the observed effect. \cite{1, 2} It is reported that light with photon energy greater than the bandgap of the semiconductor falling on a metal-semiconductor Schottky junction changes band bending at the metal-semiconductor interface and reduces the Schottky barrier height. That is claimed to results in injection of carriers from the semiconductor to the metal thereby changing work function of the metal in the illuminated part. As a consequence, if part of the device is kept in dark and part in illumination, there will be a potential difference between the part kept in dark and under illumination. \cite{1,2} The devices that generate power using this principle has been termed as shadow-effect energy generator (SEG) in the literature. \cite{1} It has been pointed out that the SEGs may be useful in self-powered photo sensing, object/motion sensing and in powering wearable electronic devices from ambient light. \cite{1}

The claim that the work function of metal electrode in Schottky junction changes under illumination is not otherwise known in literature \cite{1,2} and needs further evaluation. Specifically, it needs to be seen if alternative explanation for the SEG effect can be found that does not need to assume change in work function of the electrode upon illumination. Therefore, it is important to look for modifications in the device geometry and electrode / substrate material that not only improve device performance but also shed light on the mechanism of action and universality of energy harvesting from shadow-effect. For improved performance of the SEG device the top electrode should have two important qualities - high optical transmission in the visible and near IR region and low sheet resistance, that will enable to draw high power from the device. In these contexts, Indium tin oxide (ITO) is a possible candidate for electrode material to construct n-Si based SEGs as it makes Schottky junction with n-Si. \cite{3, 4, 5} ITO is a transparent conducting oxide having $\approx 85$ to 90\% transmission in the visible and near IR spectral region along with a resistivity of the order of $\approx 10^{-4}$ ohm-cm. \cite{6, 7, 8} In fact, ITO/n-Si heterojunction devices are already being reported as solar cell. \cite{9,10} A relatively thick layer of ITO can be used which will still have more than 80\% transmission, but have a low sheet resistance, that would enable high power to be drawn from the device. 
\begin{figure*}
\includegraphics[scale=0.6]{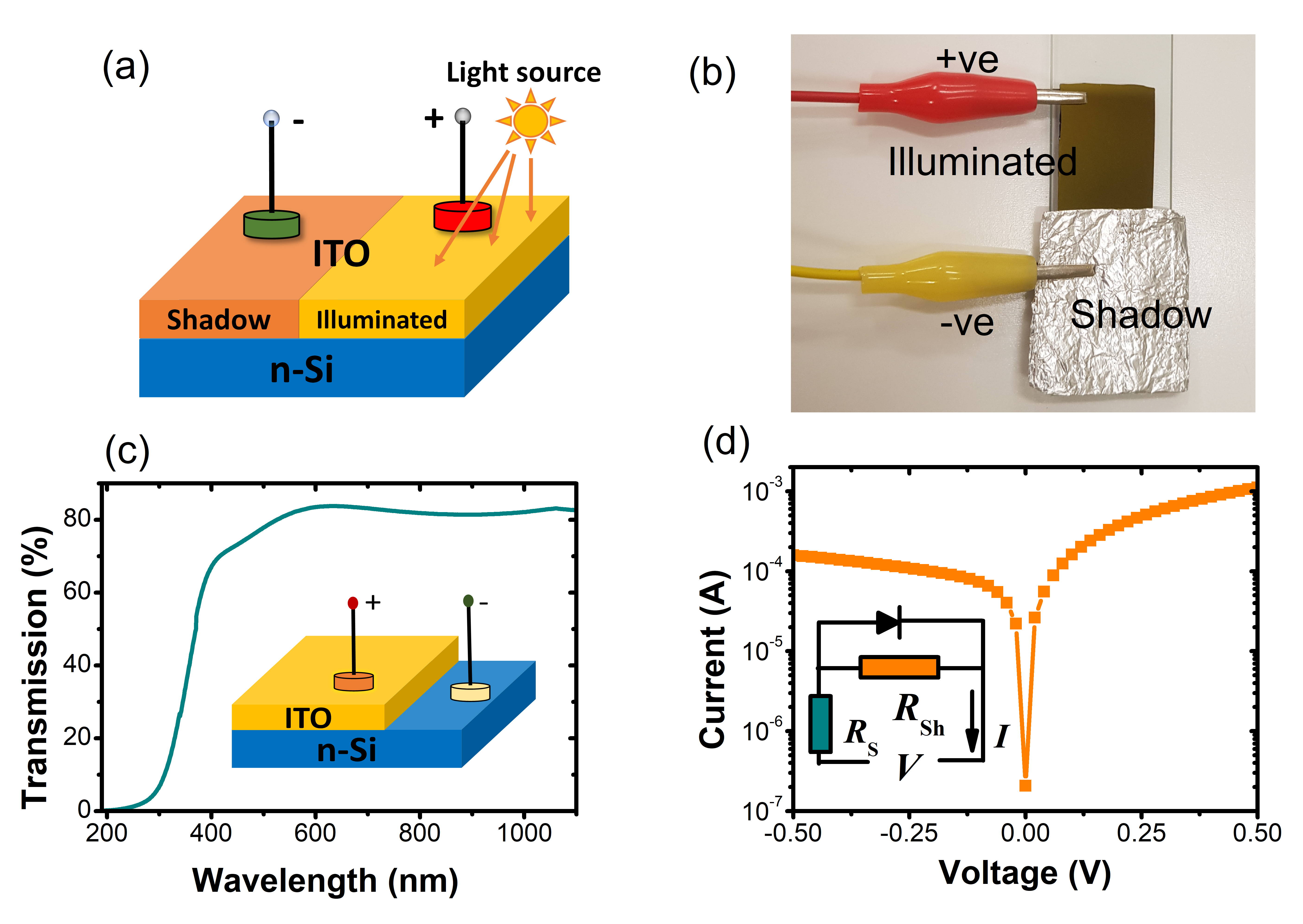}
\caption{(a) The schematic of the ITO/n-Si shadow-effect energy generator (SEG), (b) the I-V measurement scheme of the SEG. Aluminium foil is to cast shadow on one half of the device. The other half is illuminated. Crocodile clips acts as electrical connection. (c) The transmission spectra of ITO deposited on sapphire substrates. The inset shows device schematic for I-V measurement across ITO/Si junction. (d) The I-V of the SEG across the ITO/n-Si heterojunction showing Schottky characteristics in dark. The inset shows the equivalent circuit of the junction showing the series $(R_S)$ and shunt resistances $(R_{Sh})$.}
\label{fig:fig1}
\end{figure*}

In this paper, energy harvesting from shadow-effect in ITO/n-Si Schottky junctions are reported for the first time. Additionally, a novel physical mechanism to elucidate the SEG phenomenon in ITO/n-Si devices is discussed, which does not assume work function modulation of ITO electrode under illumination. Not only that, a new device geometry is also demonstrated to corroborate the proposed mechanism that enhances the open circuit voltage and the short circuit current of the SEGs as well. Moreover, the power density obtained from ITO/n-Si SEGs in both the conventional and modified geometry are higher compared to that reported in Au/n-Si SEGs.\cite{1} 

\section{Materials and methods }
The schematic of the ITO/n-Si devices, termed hereafter as SEGs for brevity, is shown in figure \ref{fig:fig1}(a). It has a simple structure wherein a thin film of ITO has been deposited on the n-Si surface. The (100) n-Si substrate (resistivity $\approx$ 1-10 ohm-cm) was cut into pieces with dimension of $\approx 4.9$ cm $\times$ 2.0 cm. The Si pieces were then cleaned using acetone and isopropyl alcohol followed by de-ionized water. The cleaned Si substrates were dipped in dilute HF (10\% v/v) solution for two minutes after which they were thoroughly rinsed in DI water and dried by nitrogen jet. The cleaned substrates were immediately loaded in the sputtering chamber. An ITO target (In¬¬2O3:SnO2 = 90:10 wt. \% with purity of 99.99 \%) was used for the sputter deposition of ITO thin film on n-Si substrates. The depositions were carried out at room temperature at RF sputtering power of 100 W and at an argon gas pressure of $7\times 10^{-3}$ mbar. The thickness of the films was measured to be $\approx 155 \pm 5$ nm. A total of four SEGs were fabricated to ascertain the device to device variability of the results. Post deposition vacuum annealing of the ITO/n-Si devices were carried out at a temperature of $200^o$C for 30 min. The annealing step was performed to improve the Schottky barrier at ITO/n-Si junction. \cite{4, 5} Moreover, ITO films were also deposited on sapphire substrates under identical deposition conditions to measure the UV-visible transmission spectra. The current-voltage measurement of the SEGs was carried out using a source-measure unit (make: Keithley). A 150 W xenon lamp (make: Oriel) was used as the light source to illuminate the device. Power of the light was measured by a white light power meter (make: Sciencetech). One typical SEG device in measurement condition is shown in figure \ref{fig:fig1}(b). Shadow was casted on one half of the device by keeping a multiply folded aluminium foil and attaching it to the device using a crocodile clip (as shown in the picture) which served as the electrical connection as well. The other crocodile clip was connected to the illuminated part of the SEG to complete the circuit.  It is emphasized here that in the SEG geometry both the electrodes connect to the ITO film, one at the illuminated part and the other at the shadowed part. The SEG had been kept on a glass plate to avoid electrical contact with backside of the Si substrate. To measure the internal photoconversion efficiency (IPCE) spectra of the SEGs, light from the Xenon lamp was passed through a $\frac{1}{4}$ meter monochromator (make: Solar TII) and made to illuminate the devices at specific wavelengths. In order to verify that ITO/n-Si junction has rectifying character, current-voltage measurement was carried out across the junction as well. For this a different device was fabricated as shown at the inset of figure \ref{fig:fig1}(c) with Mg ohmic contact on n-Si. \cite{11} The reflectance spectra and thickness of the ITO films on Si substrates were measured by using a spectroscopic reflectometer (make: Angstrom-Sun). 

\section{Results and Discussions}  
Before proceeding with the discussion of effect of illumination contrast on the ITO/n-Si SEGs, it is important to first check the transmission spectra and the sheet resistance of the ITO thin films and also to verify that Schottky junction has indeed been formed at the ITO/n-Si interface, that is essential for the performance of the SEGs. Figure \ref{fig:fig1}(c) shows the UV-visible transmission spectra of the ITO thin film deposited on sapphire substrate in the spectral range from 190 to 1100 nm. The film shows more than 80\% transmission in the spectral range from 1100 to 600 nm. The average transmission is greater than 70\% in the range between 400 to 600 nm. The transmission decays rapidly below 400 nm corresponding to the absorption edge of ITO and becomes almost zero at about 300 nm. \cite{6, 7, 8} The sheet resistance of the 155 nm thick ITO films is $\approx 45 \pm 3$ ohm. The result of current-voltage (I-V) measurement across the ITO/n-Si junction carried out in dark is shown in figure \ref{fig:fig1}(d). The I-V curve in dark shows rectifying behaviour with a rectification ratio of $\approx 7 $at 0.5 V bias. The reverse saturation current has been calculated by fitting the I-V characteristics by diode equation after correcting for the series $(R_S)$ and shunt resistances $(R_{Sh})$ as shown in the equivalent circuit of the junction at the inset in figure \ref{fig:fig1}(d). The value of reverse saturation current density ($J_s$) is calculated to be $(1.7 \pm 0.2) \times 10^{-4} $A-cm$^{-2}$ and the ideality factor is in the range from $\approx 1.1$ to 1.2. The reverse saturation current is linked to the Schottky barrier height $(\phi_B)$ by the equation $J_s = A^*T^2\exp(-\phi_B/kT)$, where $A^*$ is the Richardson's constant, $T$ is the temperature and $k$ is the Boltzmann's constant. Using the above value of reverse saturation current, the height of the Schottky barrier at ITO/n-Si junction has been estimated to be $\approx 0.643 \pm 0.004$ eV at room temperature which agrees well with work function of ITO ($\approx 4.7$ eV) and electron affinity of Si ($\approx 4.05$ eV). \cite{5, 11, 12,13} We use a value of 112 A-cm$^{-2} K^{-2}$ for the Richardson constant for n-type Si. \cite{14, 15, 16} With this background on the optical and electrical properties of ITO and ITO/n-Si Schottky junction, we proceed to study the current-voltage characteristics of the SEGs in light and dark.
\begin{figure}
\includegraphics[scale=0.36]{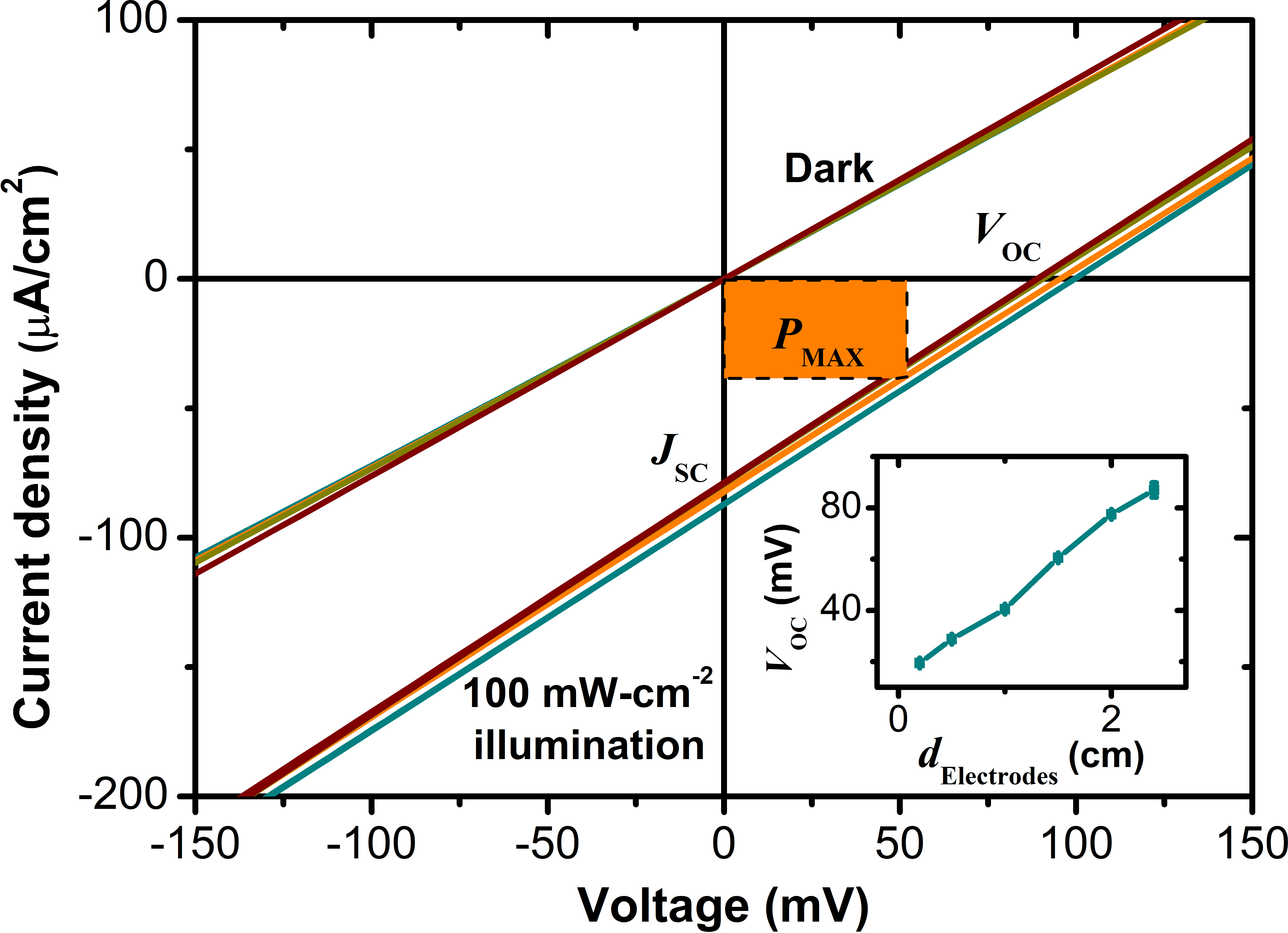}
\caption{The current density -voltage (J-V) plot of the four ITO/n-Si SEGs. The shaded area represents the maximum power $(P_{MAX})$ that can be delivered to the load. }
\label{fig:fig2}
\end{figure}

The J-V characteristics of the four different SEGs both in dark and in white light illumination are shown in figure \ref{fig:fig2}. The light intensity is $\approx$ 100 mW/cm$^2$. The J-V curve in dark is linear as expected for current flowing through ITO film. Upon light illumination, the J-V curve is still linear, but pushed downward due to generation of photovoltage. It is to be noted that only half of total surface area of the SEGs are illuminated by white light. The other half is in shadow. If the shadow region is removed, that is, the entire device is kept under uniform illumination, the J-V curve again passes through origin implying zero photovoltage. The part of the J-V curve in the fourth quadrant in figure \ref{fig:fig2} signifies power being generated by the SEGs upon receiving illumination contrast. The polarity of the voltage generated by the SEGs is such that the current flows in the external circuit from the illuminated to the shadowed part i.e. the illuminated part is at higher voltage. If the shadowed and illuminated parts are interchanged, the voltage polarity reverses. The open circuit voltage $(V_{OC})$ and the short circuit current density $(J_{SC})$ are respectively $\approx 94 \pm 5$ mV and $\approx 82 \pm 4$ $\mu$A-cm$^{-2}$ at 100 mW-cm$^{-2}$ light intensity. Although, the value of $(V_{OC})$ is lower as compared to that reported in Au/n-Si SEGs, the $(J_{SC})$ is much higher. \cite{1} It is worthwhile to mention here that we also demonstrate a method to enhance the $(V_{OC})$ of a given SEG in latter paragraphs. Maximum power density that can be delivered to the load from the SEG is equal to the area of the inscribed rectangle shown by the shaded region at the fourth quadrant of the J-V curve in the figure \ref{fig:fig2}. It corresponds to the situation when the load resistance is equal to the internal resistance of the SEGs $\it{i.e.}$ when the voltage drop across the load is half of $(V_{OC})$ and current through the load is half of the short circuit current. It is a consequence of linear I-V curve of the SEGs. The maximum power that can be delivered from an ITO/n-Si SEGs with a total device area (including the shadowed part) of $\approx 9.8\pm 0.5$ cm$^2$ to the load is $\approx 19 \pm 2$ W at 100 mW/cm$^2$ illumination. This value of the maximum power that can be delivered to load is much higher compared to that reported for Au/n-Si SEGs of similar surface area. \cite{1}

\begin{figure}[h]
\centering\includegraphics[scale=0.39]{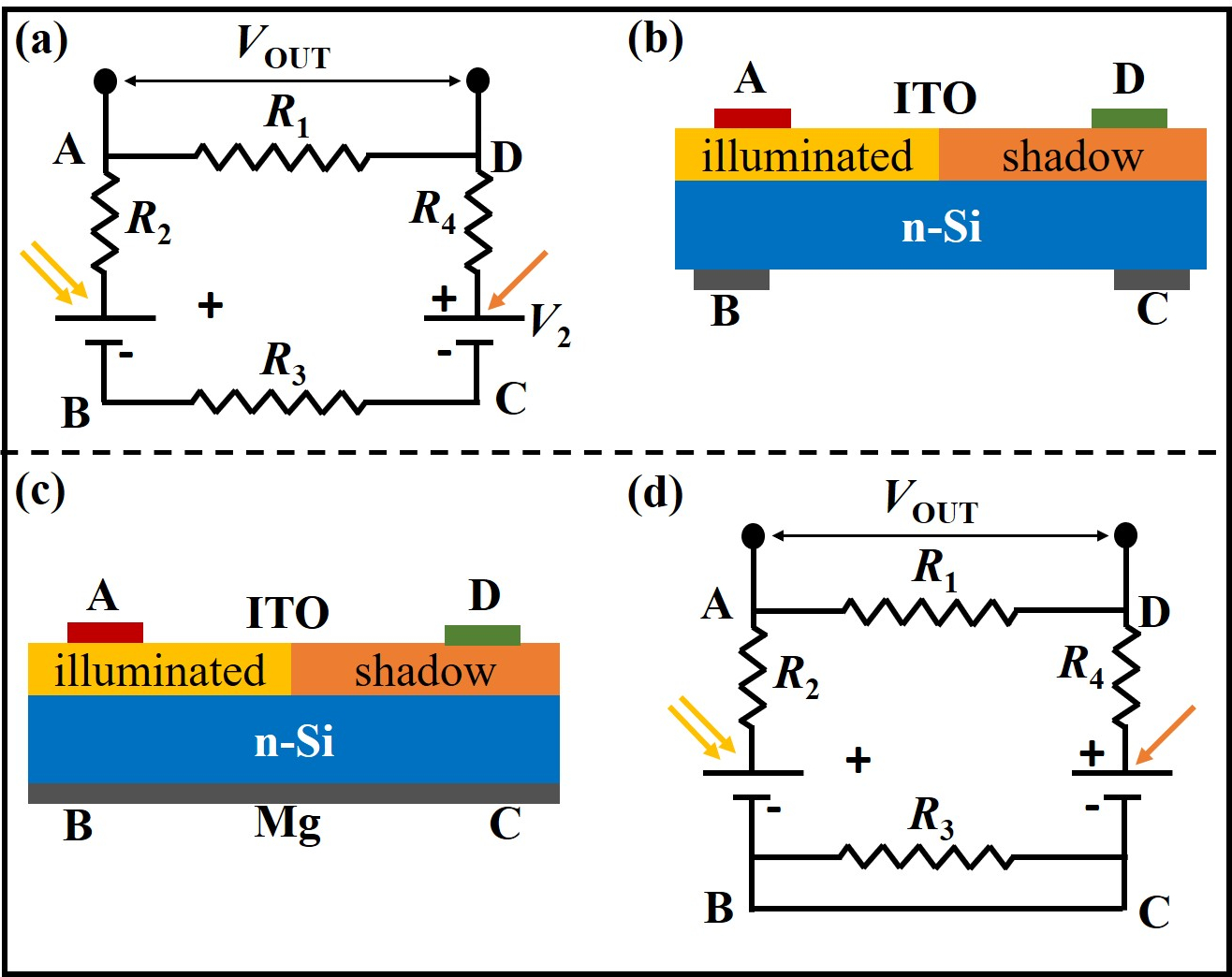}
\caption{(a) The proposed equivalent circuit of the ITO/n-Si SEG which consist of two Schottky junction solar cells connected in a closed loop by the resistances $R_1$, $R_2$, $R_3$ and $R_4$. (b)The cross section of a SEG with the electrodes A, B, C and D connected. Half of the SEGs remain in shadow and the other half is illuminated. (c) The cross section and (d) the equivalent circuit of the SEG with continuous ohmic contact at the back of n-Si substrate.}
\label{fig:fig3}
\end{figure}

\begin{figure*}
\centering\includegraphics[scale=0.6]{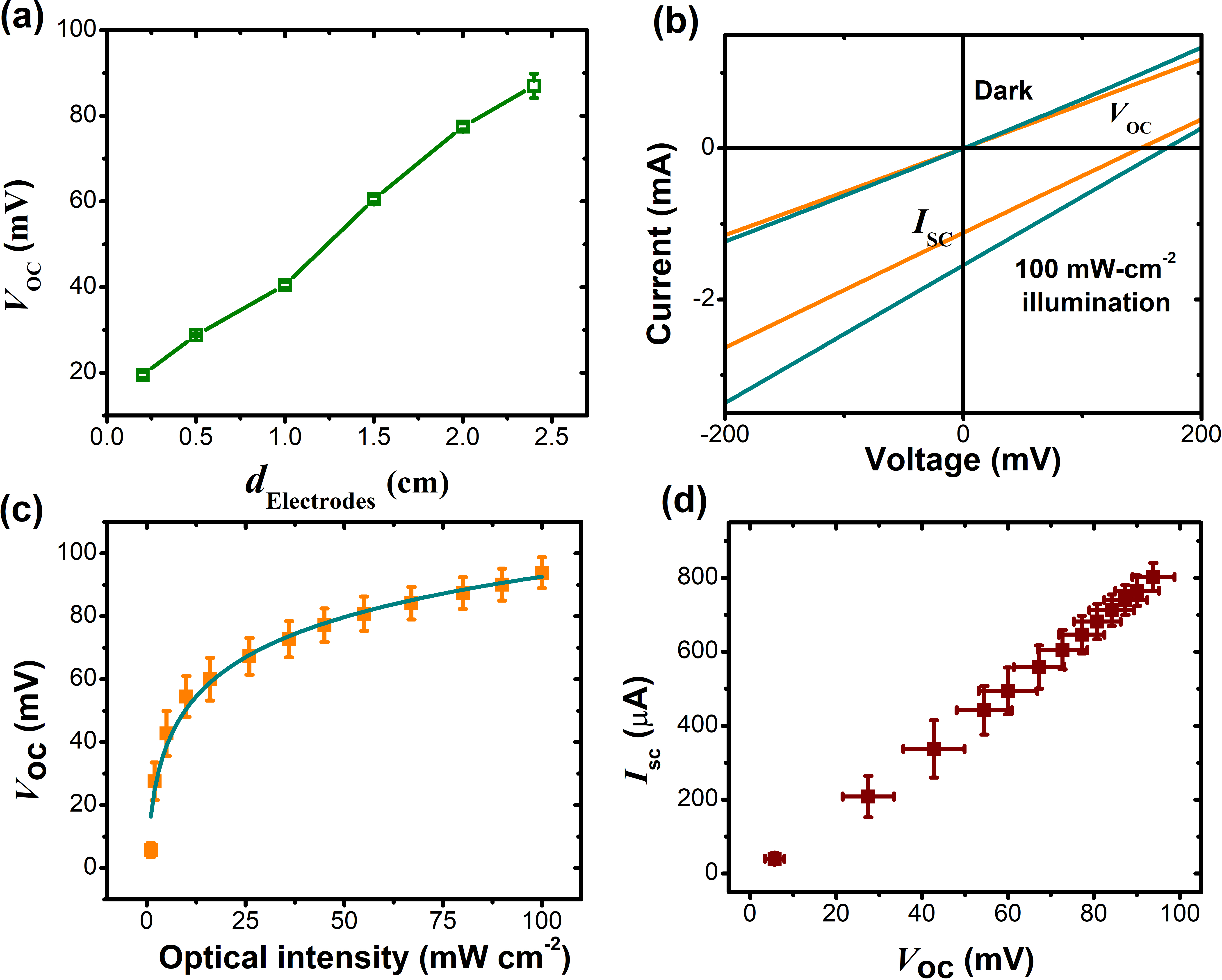}
\caption{(a) Variation of open circuit voltage ($V_{OC}$) of an SEG as a function of distance of the contact at the illuminated part from the edge of the shadowed part $(d_{Electrode})$. (b) I-V plots for the two SEGs with Mg ohmic contact at the entire back surface of the n-Si substrates in dark and in 100 mW-cm$^{-2}$ white light. (c) The variation of $V_{OC}$ (orange scattered points) of the four SEGs (prior to applying back contact) as a function of intensity of incident light. The continuous cyan curve in (b) is the fitting of the data as per equation (\ref{eq3}). (d) The short circuit current $(I_{SC})$ of the SEGs plotted as a function of the $V_{OC}$ as measured at different light intensities. }
\label{fig:fig4}
\end{figure*}

With the results of current-voltage measurement for the ITO/n-Si SEGs presented, that confirm the reported phenomenon of energy harvesting from shadow-effect, it is important now to elucidate their physical principle. As previously mentioned, the working principle of metal-semiconductor Schottky junction SEGs have been discussed by considering change in work function of metal electrode under illumination in literature. \cite{1,2} However, in the present paper we use a completely different idea that has been detailed in figure \ref{fig:fig3}, without resorting to assuming change in work function of top electrode under illumination. We propose that the SEGs are actually composed of two Schottky junction solar cells connected together in a closed loop circuit through the resistances $R_1$, $R_2$, $R_3$ and $R_4$ as shown in figure \ref{fig:fig3}(a). One of the solar cells is in shadow and the other in illumination. These two cells correspond to the respective Schottky junctions of the SEGs. The nodes A, B, C and D in figures \ref{fig:fig3}(a)-(d)  corresponds to the electrical contacts to the device. The four resistances are due to different parts of the ITO thin film and Si bulk that drop voltages. $R_1$ corresponds to the resistances of ITO film between the electrodes A and D while $R_3$ corresponds to that of the Si bulk between B and C. The resistances $R_2$ and $R_4$ are included to take into account voltage drops between the Schottky junctions and the electrodes A and D (on ITO) respectively. The inclusion of $R_2$ and $R_4$ are necessitated by the observed variation of $V_{OC}$ with distance of the contact (A) at the illuminated part from the edge of the shadowed part $(d_{Electrode})$ as depicted in figure \ref{fig:fig4}(a). As can be seen, the value of $V_{OC}$ increases with increase in $d_{Electrode}$. It is due to the fact that with increasing $d_{Electrode}$, the value of $R_1$ will increase resulting in higher voltage drop across $R_1$ ($V_{OC}$) as compared to that across $R_2$ and $R_4$. 

Refereeing to figure \ref{fig:fig3}(a), $V_{OC}$ of the SEG is the potential difference between the points A and D, which is nothing but the difference between the photovoltages of the two solar cells corrected for the voltage drops across $R_3$, $R_2$ and $R_4$. To test this idea, we measure the various potential differences across the SEGs using the device structure shown in figure \ref{fig:fig3}(b), which is a SEG with the addition of two ohmic contacts (B and C) at the n-Si substrate. The measured values of these voltages are shown at serial numbers 1 and 2 (for two of the SEGs) in Table \ref{tabone}. It is evident that the different voltages follow the equation $V_{AD} = V_{AB} + V_{BC} - V_{DC}$, that is expected for the circuit diagram in figure \ref{fig:fig3}(a). It is also interesting to note here that the magnitude of highest voltage difference is between the points A and B, implying the photovoltage being generated there ($\it{i.e.}$ at the illuminated Schottky junction) and not between A and D. Further proof of our idea can be derived by looking at the device structure of figure \ref{fig:fig3}(c), wherein an ohmic contact using Mg has been coated on the entire back surface of the n-Si substrate of the SEGs. In this geometry the voltage drop across $R_3$ can be bypassed, that is, $V_{BC}$ will be small and the net output voltage of the SEGs will be nothing but the difference between the photovoltages in the two solar cells (corrected for small voltage drops across $R_2$ and $R_4$) as depicted in the circuit diagram \ref{fig:fig3}(d). The Mg back electrode has been applied in two of the SEGs. Indeed, we find that in both the SEGs, application of the bottom ohmic layer increases $V_{OC}$ ($V_{AD}$) as shown at serial numbers 3 and 4 in Table I. The values of the various voltages in these cases also conform to the closed loop equation mentioned above. Notably, not only the open circuit voltage, but the short circuit current $(I_{SC})$ also increases on application of continuous ohmic contact at the back of n-Si substrates as can be seen from the I-V curves presented in figure \ref{fig:fig4}(b). Therefore, the proposal that the SEGs are nothing but a combination of two Schottky junction solar cells, one illuminated and the other shadowed,  in a closed loop appears to be correct. It is not at all needed to assume that the work function of ITO electrode changes under illumination to explain the phenomena. Moreover, the scheme of modifying the SEG geometry by applying an ohmic back contact to the Si substrate can be used to enhance $V_{OC}$ and $I_{SC}$ of the SEGs. Thus, higher power can be drawn from this new device geometry as compared to the conventional one.
\begin{table}
\caption{\label{tabone}Various voltages measured across different points in the SEGs with and without continuous back contact as depicted in figures \ref{fig:fig3}(b) and \ref{fig:fig3}(c). The intensity of incident white light is 100 mW-cm$^{-2}$.} 
\begin{ruledtabular}
\begin{tabular}{@{}*{6}{c}}
Serial &  Back & $V_{AB}$ & $V_{BC}$ & $V_{DC}$ & $V_{AD}$\\ 
No. &  contact &  (mV) & (mV) & (mV) & (mV)\\ 
\hline
$1$& No & $230\pm6$ & $-137\pm7$ & $7.7\pm0.9$ & $86\pm1$ \\
$2$& No & $170\pm6$ & $-64\pm7$ & $3.75\pm0.05$ & $100\pm3$ \\
$3$& Yes & $189\pm1$ & $-1.92\pm0.04$ & $18\pm1$ & $170.3\pm0.2$ \\
$4$& Yes & $154.3\pm0.7$ & $-1.44\pm0.05$ & $8.85\pm0.07$ & $144\pm3$ \\
\end{tabular}
\end{ruledtabular}
\end{table}

From the discussion above about the working principle of the SEGs it is quite clear that $V_{OC}$ of the SEGs is decided by the photovoltage generated at the two ITO/n-Si Schottky junctions. The open circuit voltage in a Schottky junction upon illumination is given as, \cite{16, 17}
\begin{equation}
V_{OC} = \left(\frac{nkT}{q}\right).\ln \left[\frac{\eta q P}{h\nu A^* T^2}\exp\left(\frac{\phi_B} {kT}\right)+1\right]
\label{eq1}
\end{equation},
where $n$ is the ideality factor, $k$ is Boltzmann's constant, $T$ is the temperature, $\eta$ is the quantum efficiency, $P$ is the intensity of light, $q$ is the electronic charge, $h$ is Plank's constant, $\nu$ is the frequency of the light, $A^*$ is the Richardson's constant. Therefore, $V_{OC}$ of the SEGs, which is a fraction of the net open circuit voltage of the two Schottky junctions, can be expressed as a function of intensity of the incident light by the following equation, 
\begin{eqnarray}
V_{OC}  & = f \left(\frac{nkT}{q}\right) (\ln \left[\frac{\eta q P_1}{h\nu A^* T^2}\exp\left(\frac{\phi_B} {kT}\right)+1\right] \nonumber\\
 & -  \ln \left[\frac{\eta q P_2}{h\nu A^* T^2}\exp\left(\frac{\phi_B} {kT}\right)+1\right])
\label{eq2m}
\end{eqnarray}
where $f$ is a number between 0 and 1. Since one of the Schottky junctions is in shadow, we can assume that the intensity $P_2$ is small compared to $P_1$ and equation (2) simplifies to,
\begin{equation}
V_{OC} = V_0 \ln \left[ aP_1 + 1\right]
\label{eq3}
\end{equation},
where constants have been subsumed in $V_0$ and $a$. The electrical performance of the ITO/n-Si SEGs as a function of intensity of the incident light is depicted in the figures \ref{fig:fig4}(c) - (d). The error bars represent sample to sample variation. The range of light intensity studied is from 1 to 100 mW-cm$^{-2}$. In the figure \ref{fig:fig4}(c) variation of $V_{OC}$ as a function of intensity of the incident light and its fitting by equation (\ref{eq3}) are shown. The scattered points are the experimental data whereas the continuous line is the fitted curve. The fit is quite good (adjusted R-squared $\approx 0.98$) which further corroborate the proposed mechanism of the SEGs. From the value of fitting parameter $a$, the quantum efficiency is calculated to be $\approx 25 \pm 6$ \% assuming $h\nu$ to be equal to the bandgap of Si. Further, referring to figure \ref{fig:fig3}(a), $I_{SC}$ can be written as $I_{SC}=V_{OC}/(R_2+R_3+R_4)$. To verify this, $I_{SC}$ of the SEGs measured at different light intensities is plotted as a function of the corresponding $V_{OC}$ in figure \ref{fig:fig4}(d). The linear nature of the resulting plot implies correctness of the above equation. It can be mentioned here as an addendum that since the two Schottky junction in the SEGs are connected in a closed loop, a current will always flow in the circuit whenever an illumination contrast is maintained, irrespective of the fact if load is present or not. This will inevitably lead to loss of energy and heating of the device.

\begin{figure}
\centering\includegraphics[scale=0.36]{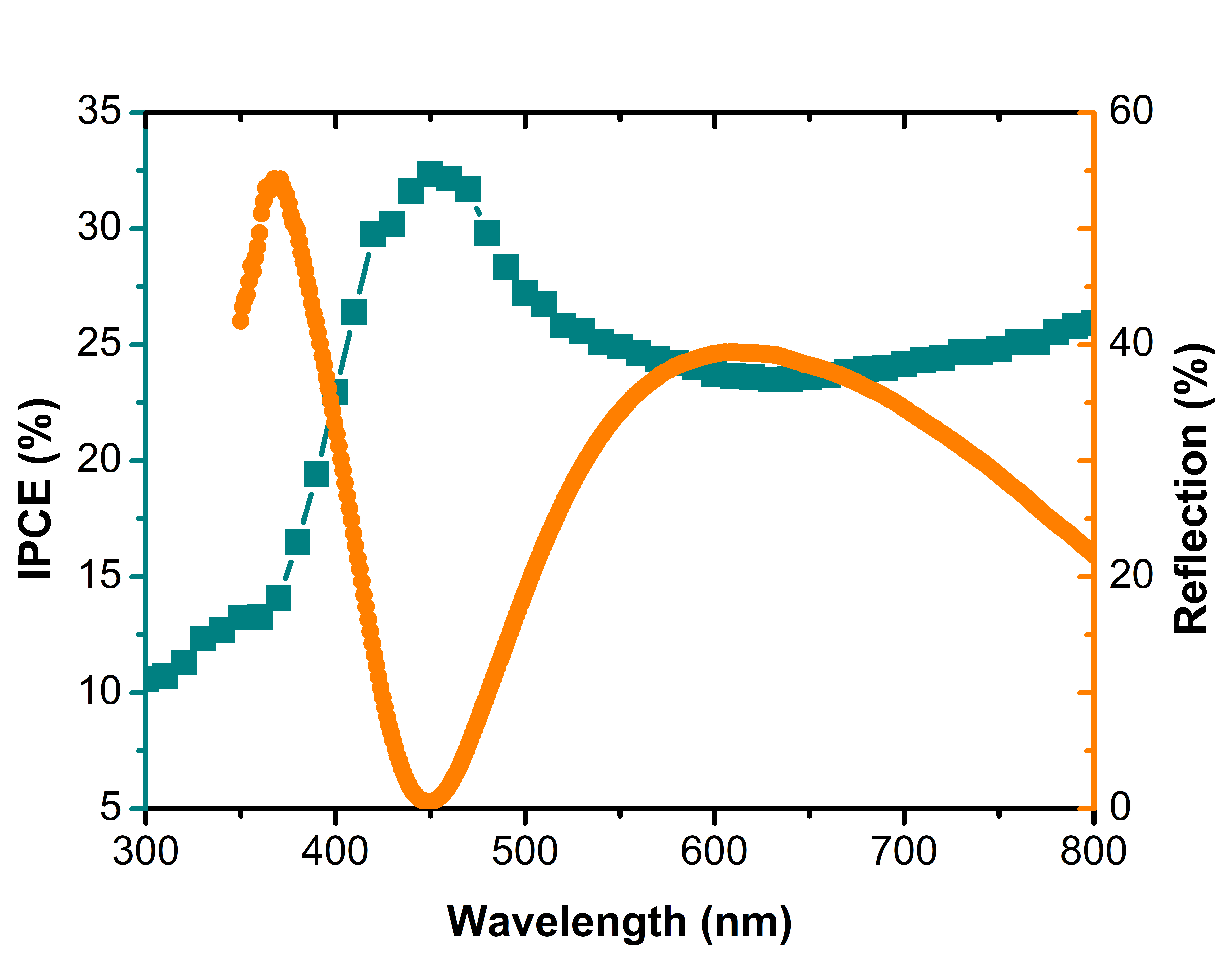}
\caption{The IPCE (dark cyan) and reflection (orange) spectra of a typical ITO/n-Si SEG. }
\label{fig:fig5}
\end{figure}

Finally, the incident photon conversion efficiency (IPCE) spectrum of the SEGs is measured to compare it to the value obtained from the fitting of experimental data as mentioned in the last paragraph. IPCE of the SEGs can be measured from a measurement of the $I_{SC}$ for a light of given wavelength and intensity as per the following equation, \cite{1, 16}
\begin{equation}
IPCE(\%) = \frac{1240\times I_{SC} (A-cm^{-2})\times 100}{\lambda (nm) P (W-cm^{-2})}
\label{eq4}
\end{equation}
The typical IPCE spectra of an ITO/n-Si SEG in the spectral range from 300 to 800 nm is shown in figure \ref{fig:fig5}. The peak value of IPCE is $\approx 32$\% at 450 nm. IPCE curve falls rapidly for wavelengths less than 400 nm implying absorption of light by ITO film. For wavelength ranging from 500- 800 nm, the value of IPCE hover around 25\%, which is close to the value obtained from the fitting as per equation (\ref{eq3}) (figure \ref{fig:fig4}(b)), signifying the appropriateness of the equation. Figure \ref{fig:fig5} also depicts the reflectance spectra of the device. It is interesting to note that the IPCE spectra has almost one-to-one correspondence with the reflection spectra. It is expected as reflection loss of light hampers the incident photon conversion efficiency. Therefore, to gain better efficiency from the ITO/n-Si SEGs, design has to be optimized for reducing the reflection losses.

\section{Conclusion} 
In conclusion, the recently invented phenomenon of energy harvesting from shadow-effect has been observed in ITO/n-Si Schottky junction that produces higher power density as compared to the Au/n-Si based devices reported in literature. A novel physical mechanism of the observed effects is presented considering the shadow-effect energy generator device as a combination of two Schottky junction solar cells, one illuminated and the other shadowed, connected in a closed loop circuit. This mechanism does not rely on considering modulation of work function of the top electrode of the junction under illumination as has been reported in literature so far. The observed data agree well with the proposed mechanism. A new device structure for shadow-effect energy generator is also demonstrated based on the proposed mechanism that enhances both the open circuit voltage as well as short circuit current. Since the equivalent circuit of the shadow-effect generators consist of two Schottky junctions connected in a closed loop circuit, a current will always be flowing through the device under contrasting illumination even if a load is not present, which will lead to wastage of harvested energy. Nevertheless, the shadow-effect energy generators can be used for self-powered photo sensing, object/motion detection and also for producing power from contrasting illumination. 

\begin{acknowledgments}
Authors acknowledge Shri Rakesh Kaul, Head, Laser Material Processing Division, Raja Ramanna Centre for Advanced Technology, Indore for the constant support and encouragement during this work.
\end{acknowledgments}


\end{document}